\documentclass[prb,twocolumn,superscriptaddress]{revtex4-1}

\usepackage{amsmath}
\usepackage{amssymb}
\usepackage{xspace}
\usepackage{graphicx}
\usepackage{grffile}
\usepackage{nicefrac}
\usepackage{color}

\newcommand{\alphabar}{\bar{\alpha}}
\newcommand{\betabar}{\bar{\beta}}
\newcommand{\gammabar}{\bar{\alpha}'}
\newcommand{\Abar}{\bar{A}}

\newcommand{\Cbar}{\bar{A'}}
\newcommand{\abs}[1]{\lvert #1 \rvert}
\newcommand{\mexc}{\stackrel{\sim}{\mathbf{m}}}

\newcommand{\cop}[1]{c_{#1}^{\phantom{\dagger}}}
\newcommand{\cdaggop}[1]{c_{#1}^{\dagger}}
\begin{document}

\title{Versatile Approach to the Spin Dynamics in Correlated Electron Systems}

\author{Malte Behrmann}
\affiliation{I. Institut f{\"u}r Theoretische Physik, Universit{\"a}t Hamburg, 
D-20355 Hamburg, Germany}
\author{Alexander I. Lichtenstein}
\affiliation{I. Institut f{\"u}r Theoretische Physik, Universit{\"a}t Hamburg, 
D-20355 Hamburg, Germany}
\author{Mikhail I. Katsnelson}
\affiliation{Radboud University Nijmegen, Institute for Molecules and Materials,
NL-6525 AJ Nijmegen, The Netherlands}
\author{Frank Lechermann}
\affiliation{I. Institut f{\"u}r Theoretische Physik, Universit{\"a}t Hamburg, 
D-20355 Hamburg, Germany}
\affiliation{Institut f{\"u}r Keramische Hochleistungswerkstoffe, Technische Universit\"at
Hamburg-Harburg, D-21073 Hamburg, Germany}


\begin{abstract}
Time-dependent spin phenomena in condensed matter are most often either described in the
weakly correlated limit of metallic Stoner/Slater-like magnetism via band theory or in the
strongly correlated limit of Heisenberg-like interacting spins in an insulator. However many
experimental studies, e.g. of (de)magnetization processes, focus on itinerant local-moment
materials such as transition metals and various of their compounds. We here present a
general theoretical framework that is capable of addressing correlated spin dynamics, also
in the presence of a vanishing charge gap. A real-space implementation of the time-dependent
rotational-invariant slave boson methodology allows to treat non-equilibrium spins numerically fast 
and efficiently beyond linear response as well as beyond the band-theoretical or Heisenberg 
limit.
\end{abstract}

\maketitle

\section{Introduction}
Non-equilibrium physics in challenging condensed matter electron systems, 
gained enormous attention in recent times. Highlight studies include e.g. driven 
(de)magnetization in transition-metal~\cite{bea96} or rare-earth~\cite{mel08} systems, 
light-induced superconductivity~\cite{fau11} or melting of charge-density 
waves.~\cite{schm08,hel12} Introducing explicit time dependence within
interacting quantum materials could allow for a dynamic stabilization of equilibrium
metastable states. Furthermore new pathways to novel exotic states
of matter may thereby envisioned.

Dynamic magnetism is a key research focus in this respect, since it not only continues 
the history of longstanding studies of an ubiquitous solid-state phenomenon. There is also
always the chance for groundbreaking technological applications. While experimental 
progress has been fast and investigations nowadays deal with a wide range of materials, 
the theoretical description struggles to keep up. So far only two limiting regimes are 
reasonably well accessible by theoretical means. First the metallic band-magnetism 
limit without the notion of local-moment physics, and second the insulating 
pure-spin limit where charge degrees of freedom are gapped. For the former case there are 
several modeling ideas within band theory dealing with itinerant Stoner physics out of 
equilibrium. Elliot-Yaffet(-like) theory~\cite{ell54,yaf63} e.g. is a widely 
utilized~\cite{steiauf_2009,kraus_2009} theoretical framework to address metallic 
demagnetization experiments. Model Hamiltonians of Heisenberg kind are on the contrary often 
applied to the interacting problem of time-dependent localized lattice spins.~\cite{ost12}
Yet most concrete experimental studies focus on itinerant systems with coexisting local 
moments and/or involve some charge fluctuations in the dynamic-probing protocol. In fact, 
it is agreed that the interplay of band theory and electron correlation is at the heart of 
dynamic materials magnetism.~\cite{see15,chi15,say15} Hence theory should be ready to tackle the 
generic problem of non-equilibrium magnetic phenomena aside from the pure-band and 
-spin limits.   

Time-dependent (TD) correlation phenomena are describable in a close-to-exact numerical manner
within TD density-matrix-renormalization-group (DMRG) approaches,~\cite{whi04,dal04} however 
these techniques are so far restricted to lattice problems in one spatial dimension.
Correlated magnetism on higher-dimensional lattices starts to be 
investigated~\cite{werner_2012,men14} by time-dependent dynamical 
mean-field theory (TD-DMFT).~\cite{schmi02,fre06,eckstein_2009} But the Keldysh-based
method is numerically very heavy and not yet capable to address general problems dealing with 
an interplay between doping, metallicity, non-collinearity, inhomogeneous features 
and/or multi-orbital degrees of freedom. There are simpler e.g. Gutzwiller-based 
non-equilibrium schemes.~\cite{schiro_2010} In view of concrete spin-dynamics problems those 
however are so far restricted to the linear-response 
limit~\cite{sei04,oelsen_2011,bunemann_2013} or are put into practise within idealized 
model settings.~\cite{sandri_2013} 

The aim of the present work is to introduce a novel approach to the spin dynamics 
emerging from correlated electrons, without restriction to too advantageous theory limits.
Our real-space implementation of the time-dependent rotational-invariant auxiliary
(or 'slave')-boson (TD-RISB)
scheme opens the possibility to study general interacting problems out of equilibrium in 
an efficient and flexible way. This is here demonstrated by its reliability in mediating
between the linear-response limit of spin excitations within the Slater and the Heisenberg 
limits of the Hubbard model. Global and local excitations beyond linear response in the doped 
Mott-insulating regime exhibit the vast potential of the versatile framework.

The RISB approach to equilibrium problems of multi-orbital correlated electrons has been 
proven successful for model Hamiltonians~\cite{li89,lec07,fer09,isi09,schu12,beh15,fac16} as 
well as in the context of realistic materials.~\cite{lec09,maz14} In essence, the method may be 
characterized on the operator level by the decomposition of the complete electron degree of freedom 
$c^{(\dagger)}$ into a low-energy quasiparticle (fermionic) part $f^{(\dagger)}$ and high-energy 
Hubbard (bosonic) representants $\{\phi^{(\dagger)}\}$. In a key approximation, the bosonic 
degrees of freedom are treated
on the mean-field level, hence the simplified electron self-energy is, as in more
general DMFT, purely local. Extension to the time domain, motivated by previous Gutzwiller 
advances,~\cite{schiro_2010} opens the possibility for the description of intricate 
multi-orbital electon correlations out of equilibrium.~\cite{beh13,beh15} Time propagation 
of quasiparticle (QP) and bosonic degrees of freedom is described by a set of coupled 
non-linear Schr\"odinger-like equations. Here the TD version is implemented in real space, 
which opens the possibility to study inhomogeneous lattice dynamics as well as two-particle 
excitations beyond linear response. 

\section{Model and Methodology}
\subsection{Interacting Hamiltonian}
\begin{figure}[t]
\includegraphics*[width=7cm]{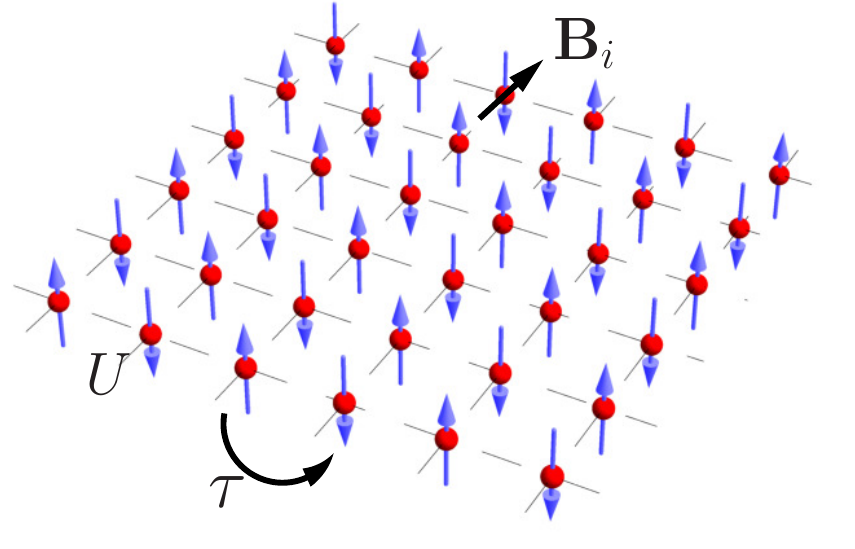}
\caption{(color online) Hubbard model with applied site-dependent magnetic field on a
$6\times 6$ real-space lattice.}\label{fig:lattice}
\end{figure}

We focus on the one-band Hubbard model on a square lattice with hopping 
$\tau$, on-site interaction $U$ as well as a space- and time-dependent magnetic 
field $B({\bf r},t)$, i.e.
\begin{equation}
{\cal H}(t) = -\sum_{\ij\sigma}\tau_{ij}^{\hfill}\, \cdaggop{i \sigma}\cop{j \sigma}\, 
-\sum_i \mathbf{B}_i (t) \cdot \mathbf{S}_i +
U\, \sum_{i} n_{i \uparrow} n_{i \downarrow}\;, \label{eq:dyn_ham}
\end{equation}
where $i,j$ label lattice sites, $\sigma=\uparrow,\downarrow$ marks the spin projection 
and $\mathbf{S}$ is the local spin operator. A real-space lattice of size $N=6\times 6$ 
(cf. Fig.~\ref{fig:lattice}) with periodic boundary conditions is employed. Extension to 
multi-orbital problems with a concrete materials background is 
straightforward.~\cite{beh13,beh15_2}

\subsection{Rotational-invariant slave-boson (RISB) representation}
To introduce our framework, lets first discuss the equilibrium case at $t=0$ and remind of the
state-of-the-art RISB technique. For more details and generalizations we refer 
to Ref.~\onlinecite{lec07}.

In real space, the one-band limit of the method asks for a set
\begin{equation}
\mbox{site}\;i=1,N\,:\;\; f_\uparrow\;,\;f_\downarrow\;\;;\;\;\phi_E^{\hfill}\;,\;
\left(\begin{array}{cc}
\phi_{\uparrow\uparrow} & \phi_{\uparrow\downarrow} \\
\phi_{\downarrow\uparrow} & \phi_{\downarrow\downarrow} \end{array}\right)\;,\;
\phi_D^{\hfill}
\end{equation}
of degrees of freedom to provide a complete state representation, whereby 
$\phi_E^{\hfill},\,\phi_D^{\hfill}$ 
are associated with the empty and doubly-occupied site. A full coverage of the spin-rotational 
latitude is given by the slave-boson matrix for the singly-occupied site, and the whole set of 
bosons still accounts for possible local charge fluctuations.
The four possible electron states 
$\{A\}=\{|E\rangle, |\uparrow\rangle, |\downarrow , |D\rangle\}$ 
on a single lattice site are represented in RISB upon action on the vauum 
state $|{\rm vac}\rangle$ as follows
\begin{eqnarray}
|E\rangle=|0\rangle &=&\phi_E^{\dagger}\,|{\rm vac}\rangle\\
|\uparrow\rangle &=& 
\frac{1}{2}\left\{\phi_{\uparrow\uparrow}^{\dagger}\,f_{\uparrow}^{\dagger}+
\phi_{\uparrow\downarrow}^{\hfill}\,f_{\downarrow}^{\dagger}\right\}\,|{\rm vac}\rangle\\
|\downarrow\rangle &=& 
\frac{1}{2}\left\{\phi_{\downarrow\uparrow}^{\dagger}\,f_{\uparrow}^{\dagger}+
\phi_{\downarrow\downarrow}^{\hfill}\,f_{\downarrow}^{\dagger}\right\}\,|{\rm vac}\rangle\\
|D\rangle =|\uparrow\downarrow\rangle&=&\phi_D^{\dagger}\,|{\rm vac}\rangle\quad.
\end{eqnarray}
The second index on the single-particle bosons refer to a QP degree of freedom, whereas the 
first index generally is associated with the local state. In order to select
the true physical states, the constraints
\begin{eqnarray}
1&=&\phi_E^\dagger\phi_E^{\hfill}+\sum_{\sigma\sigma'}
\phi^\dagger_{\sigma\sigma'}\phi_{\sigma\sigma'}^{\hfill}+
\phi^\dagger_D\phi_D^{\hfill}\label{eq:con1}\\
f^\dagger_\sigma f_\sigma^{\hfill}&=&\phi_D^\dagger\phi_D^{\hfill}
+\sum_{\sigma'}\phi^\dagger_{\sigma\sigma'}\phi_{\sigma\sigma'}^{\hfill}\label{eq:con2}\\
f^\dagger_{\sigma}f_{\bar{\sigma}}^{\hfill}&=&
\sum_{\sigma'}\phi^\dagger_{\sigma'\bar{\sigma}}\phi_{\sigma'\sigma}^{\hfill}\label{eq:con3}
\end{eqnarray}
have to be enforced on each site. We then write the interacting Hamiltonian (\ref{eq:dyn_ham}) 
in equilibrium as 
$\underline{\cal H}=\underline{\cal H}^{\rm (kin)}+\sum_i\underline{\cal H}^{\rm (loc)}_i$, 
whereby the electron operator is expressed through
\begin{eqnarray}
\underline{c}_{i\sigma}^\dagger&=&\frac{1}{\sqrt{2}}
\sum_{\sigma'}\left\{\phi_{i\sigma\sigma'}^\dagger\phi_{iE}^{\hfill}-
(-1)^{\delta_{\sigma\sigma'}}\phi^\dagger_{iD}\phi_{i\bar{\sigma}\bar{\sigma}'}^{\hfill}\right\}
\,f_{i\sigma'}^\dagger\nonumber \\
&\equiv&\sum_{\sigma'} R_{i\sigma'\sigma}^{\dagger}\,f_{i\sigma'}^\dagger\quad.\label{eq:cop}
\end{eqnarray}
Note that as common in slave-particle theories, there is a gauge symmetry providing some freedom 
in the actual representation of the QP indices on each lattice site. But as shown in 
Ref.~\onlinecite{lec07}, physical observables remain of course gauge invariant. Let us mention 
that in this regard, Lanata et al.~\cite{lan16} recently proposed an alternative RISB 
representation.

We can write the kinetic Hamiltonian readily as
\begin{equation}
\underline{\cal H}^{\rm (kin)}=\sum_{ij}\sum_{\sigma\sigma'\sigma''}
R_{i\sigma'\sigma}^{\dagger}\,\tau_{ij}^{\hfill}\,R_{j\sigma\sigma''}^{\hfill}
\,f_{i\sigma'}^\dagger f_{j\sigma''}^{\hfill}\quad.\label{eq:hkin}
\end{equation}
One may define a local QP weight via 
${\bf Z}_{i}^{\hfill}={\bf R}_i^{\hfill}{\bf R}^{\dagger}_i$. 
Eqns.~(\ref{eq:cop},\ref{eq:hkin}) already render the key feature for
describing non-trivial spin dynamics obvious: the rotational-invariant framework allows for 
spin $\sigma,\sigma'$ exchange through local-multiplet excitations via QP hopping 
processes. 

To represent the local Hamiltonian, one uses the fact that any local operator ${\cal O}$ 
may be written in quadratic terms of the bosonic degrees of freedom. The general RISB 
form is given by
\begin{equation}
\underline{\cal O}=\sum_{AA'}
\langle A|{\cal O}|A'\rangle\sum_{\gamma}\phi^\dagger_{A\gamma}\phi^{\hfill}_{A'\gamma}\quad.
\end{equation}
For the local Hubbard interaction, i.e. 
$\underline{\cal O}=U\underline{n}_{i \uparrow} \underline{n}_{i \downarrow}$,
the slave-boson representation ${\cal H}_{\rm U}=U\phi_D^\dagger\phi_D^{\hfill}$ is readily 
obtained. With the help of the Pauli matrices \boldmath${\cal S}$\unboldmath$_\nu$ along 
the component $\nu=x,y,z$, the local spin operator generally reads 
${\bf S}=\frac{1}{2}\sum_{\sigma\sigma'}
c_{\sigma}^\dagger\,$\boldmath$\vec{{\cal S}}$\unboldmath$\,c_{\sigma'}^{\hfill}$. In the
Hamiltonian~(\ref{eq:dyn_ham}) it appears in the product form 
${\bf B}\cdot{\bf S}=\sum_\nu B_\nu\,S_\nu$. Hence,
\begin{eqnarray}
\underline{{\bf B}\cdot{\bf S}}&=&
\sum_{AA'}\langle A|{\bf B}\cdot{\bf S}|
A'\rangle\sum_{\gamma}\phi^\dagger_{A\gamma}\phi^{\hfill}_{A'\gamma}\nonumber\\
&=&\frac{1}{2}\sum_\nu B_\nu\sum_{\sigma\sigma'}\langle\sigma|
S_\nu|\sigma'\rangle\sum_{\sigma''}
\phi^\dagger_{\sigma\sigma''}\phi_{\sigma'\sigma''}^{\hfill}\nonumber\\
&=&\frac{1}{2}\sum_\nu B_\nu\sum_{\sigma\sigma'}{\cal S}_{\nu\sigma\sigma'}
\sum_{\sigma''}\phi^\dagger_{\sigma\sigma''}\phi_{\sigma'\sigma''}^{\hfill}\quad.
\end{eqnarray}
Together with the Hubbard interaction, this completes the local-Hamiltonian representation
\begin{eqnarray}
\underline{\cal H}_i^{\rm (loc)}&=&-\frac{1}{2}\sum_{\nu=xyz}B_{i\nu}^{\hfill}
\sum_{\sigma\sigma'\sigma''}{\cal S}_{\nu\sigma\sigma'}\,
\phi^\dagger_{i\sigma\sigma''}\phi_{i\sigma'\sigma''}^{\hfill}+\nonumber\\
&&+\;U\,\phi^\dagger_{iD}\phi_{iD}^{\hfill}\quad.
\end{eqnarray}
For the rest of the paper, the hopping $\tau$ is restricted to nearest neigbours. 
Energies(Times) will be given in units of the (inverse) half-bandwidth $\frac{W}{2}=4\tau$.

In the following, the mean-field limit of the RISB theory is put into practise. Within that
limit (i.e. at saddle-point) the bosonic degrees of freedom are condensed and treated as $c$-numbers. 
Proper normalizations of the respective RISB electronic operators have the to be invoked to ensure a 
coherent description in the different interaction limits. This means that the $R$-matrices introduced 
in eq. (\ref{eq:cop}) are normalized such to yield the correct QP-weight limit $Z$=1 at weak coupling. 
For a detailed discussion of this matter we refer to Ref.~\onlinecite{lec07}. Note that it has been 
shown~\cite{bue07} that the mean-field RISB method is equivalent to the generalized multi-orbital 
Gutzwiller approach.~\cite{bue98}

\subsection{AFM ground state at half filling}
For our considerations, the initial phase at $t=0$ is defined by the interacting lattice in 
equilibrium. It is given by the symmetry-broken antiferromagnetic (AFM) ordered 
phase close to half filling, where the local moments are collinear initialized in $x$-direction.
Some key physical quantities of the equilibrium AFM lattice at half filling as described
in RISB are given in Tab.~\ref{tab:rs-dyn_AFM_eq_results}. 

At weak coupling with $U=0.4$, a band-magnetic Slater limit is observed, in which the local 
magnetic response is comparatively small. Increasing $U$ enhances the local and ordered 
moments, until at $U=3$ the spins are almost fully polarized. The time-independent form of the 
Hubbard Hamiltonian (\ref{eq:dyn_ham}) then starts to approach a Heisenberg-Hamiltonian form
which is fully realized at very large $U$. Note that for all values of $U$ a finite charge 
gap persists, i.e. the equilibrium ground state is an AFM insulator. Since the paramagnetic 
Mott transition in the present scheme sets in at $U\sim 3.05$, the value of $U=3$ is located 
in the strongly correlated regime of the Hubbard model.
\begin{table}[t]
\centering
\begin{tabular}{ c | c | c | c | c  }
Lattice size & $U$ & $\abs{\left<m\right>}\le 1.0$ & $\left<S^2\right>\le 3/4$ & charge gap \\
\hline
6$\times$6 & 0.4 & 0.39 & 0.46  & 0.06 \\
& 1.0 & 0.62 & 0.56  & 0.33 \\
& 2.0 & 0.86 & 0.67  & 2.40 \\
& 3.0 & 0.94 & 0.71  & 3.22 \\
\hline  
8$\times$8 & 0.4 & 0.34 & 0.45 & 0.06 \\
\end{tabular}
\caption{Physical quantities extracted from the homogeneous equilibrium AFM state for different 
Hubbard $U$: ordered magnetic moment $\langle m\rangle$, local spin moment $\langle S^2\rangle$ 
and charge gap.}\label{tab:rs-dyn_AFM_eq_results}
\end{table} 

\subsection{Time-dependent RISB scheme}
For the dynamic regime, the equilibrium RISB solution sets 
the stage. We then approximately propagate the equilibrium solution with the TD Hamiltonian 
(\ref{eq:dyn_ham}). The condensed slave bosons $\phi$ become time dependent and a set of
non-linear differential equations governs the problem. 
Conveniently, a multi-index notation is used, whereby a site-dependent quantity $a_i$ can be 
written as $\bar{a}=\left(i,a\right)$. Greek letters label the QP degrees of 
freedom and $A,A'$ describe local states. Then the set of differential equations 
reads~\cite{beh16,fab13} 
 \begin{eqnarray}
 i\,\frac{\partial \eta_{\betabar \alphabar}^{\hfill}}{\partial t} &=& 
\sum_{\bar{\beta}'} \underline{\cal H}^{\rm (kin)}_{\alphabar \bar{\beta}'} 
\eta_{\betabar \bar{\beta}'}^{\hfill} \label{td_dgl1}\\
 i\,\frac{\partial \phi_{\Abar \bar{\gamma}}^{\hfill}}{\partial t} &=& \sum_{\Cbar} 
\underline{\cal H}^{\rm (loc)}_{\Abar \Cbar} 
\phi_{\Cbar \bar{\gamma}}^{\hfill} +\sum_{\gammabar}^{\rm occ}
\sum_{\alphabar \betabar} \eta_{\alphabar \gammabar}^{\dagger} 
\frac{\partial \underline{\cal H}^{\rm (kin)}_{\alphabar \betabar}}{\partial 
\phi_{\Abar\bar{\gamma}}^\dagger}\eta_{\gammabar \betabar}^{\hfill} \label{td_dgl2}\quad.
 \end{eqnarray}
In the first subset (\ref{td_dgl1}) of equations, $\betabar$ runs only over occupied sites and
spin projections, while $\alphabar,\bar{\beta}'$ run over all sites and
spin projections.  The quantity $\eta$ depicts eigenstates of the renormalized kinetic Hamiltonian
$\underline{\cal H}^{\rm (kin)}$ in real space. The restriction to occupied states is indicated 
in the second equation subset (\ref{td_dgl2}). It means that in TD-RISB (as in TD-Gutzwiller)
the fixed quasiparticle occupations enter the time evolution of the bosonic (or 
Gutzwiller-projector) degrees of freedom. Still, note that the character of the QP states is  
time dependent and non-equilibrium charge fluctuations/transfers of the physical electrons
are allowed. 
A numerical solution of eqns.~(\ref{td_dgl1},\ref{td_dgl2}) is achieved by using an adaptive 
Runge-Kutta scheme of sixth-fifth order~\cite{verner_numerically_2010}. 
In the TD scheme, the constraints (\ref{eq:con1}-\ref{eq:con3}) are established at 
$t=0$ and then remain fulfilled during the time development, i.e. the constraints are integrals
of motion.

A specific excitation of the lattice induced by the magnetic-field term in the Hamiltonian 
(\ref{eq:dyn_ham}) is performed to achieve two goals. First, we want to measure  
the magnetic excitations with a single TD calculation. Second aim is to efficiently scan the
parameter space in order to tune the response regime from the linear-response limit to the fully 
non-equilibrium domain. The linear-response regime is mainly investigated to compare to previous 
studies and to establish a basic understanding, while the latter provides new insight in fully TD
magnetic excitations. 

The necessary mean-field normalizations of the RISB operators, governing also the
TD equations~(\ref{td_dgl1},\ref{td_dgl2}), are identical to the ones at equilibrium and
are properly included in the calculations. To illustrate the reliability of our time-dependent
scheme, we provide in the appendix results for the canonical TD one-band Mott transition after 
an interaction quench, in perfect agreement with TD-Gutzwiller 
calculations.~\cite{schiro_2010}

\section{Spin Dynamics}
We first describe in section~\ref{pulse} the modeling of the excitation of the lattice system as well
as the monitoring of the subsequent time evolution. In the following sections~\ref{slahe}-\ref{locx} 
different concrete applications are addressed and discussed.

\subsection{Pulsed excitation\label{pulse}}
\begin{figure}[t]
\centering
\includegraphics[width=8cm]{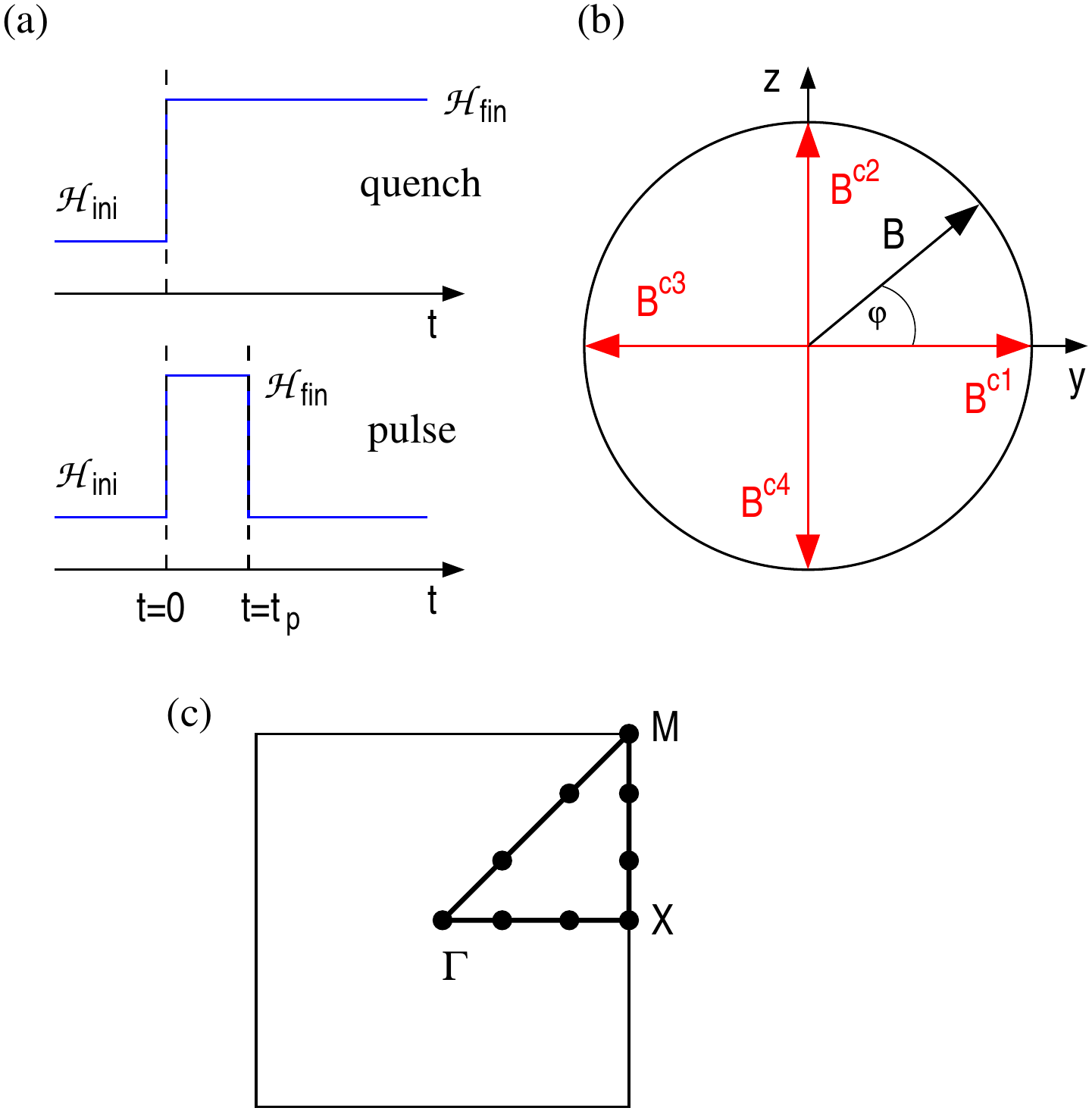}
\caption{On the time-dependent calculational settings. 
(a) Difference between a quench and a pulse with duration time $t_p$. 
(b) Sketch of the possible $C=4$ magnetic-field configurations, applied during $t_p$ via
$\mathcal{H}_{\rm fin}$ at lattice site $i$.
(d) Brillouin-zone path along the high-symmetry points $\Gamma=(0,0)$, $X=(\pi,0)$ 
and $M=(\pi,\pi)$ of the square lattice. Dots mark the accessible points 
$\mathbf{q}=(q_y,q_z)$ for the 6$\times$6 lattice.}
\label{fig:rs-dyn_angle_kick_quench}
\end{figure}

Instead of quenching the system, the TD Hamiltonian is applied in a pulsed form (see 
Fig.~\ref{fig:rs-dyn_angle_kick_quench}a). In a quench scenario, there is a sudden switch 
of the initial equilibrium Hamiltonian $\mathcal{H}_{\rm ini}=\mathcal{H}(t$$=$$0)$ to a final 
Hamiltonian $\mathcal{H}_{\rm fin}=\mathcal{H}(t$$>$$0)$. On the contrary in the pulsed case, the
final-Hamiltonian form $\mathcal{H}_{\rm fin}=\mathcal{H}(0$$<$$t$$<$$t_p)$ holds, whereby $t_p$ 
marks the duration of the magnetic pulse. After time $t_p$ the Hamiltonian is 
switched back to $\mathcal{H}_{\rm ini}$. Note that both, the general time $t$ and the 
pulse-duration time $t_p$ have the identical zero value.

The site-dependent magnetic field $\mathbf{B}(\mathbf{r},t)=\mathbf{B}_{i}(t)$ in 
$\mathcal{H}_{\rm fin}$ is here chosen as random in the $yz$-plane with zero $x$-component 
(see Fig.~\ref{fig:rs-dyn_angle_kick_quench}b). The fixed absolute value 
is site independent, i.e. $\abs{\mathbf{B}_{i}}=B$. Since we deal with a 
finite lattice resolution, it is beneficial to consider only a discrete number of inhomogeneous 
magnetic-field configurations $\mathbf{B}_{i}$ in the $yz$-plane. 
We consider the angle $\varphi$ to the $y$-axis and the discrete configurations 
$\varphi=\frac{2\pi\,\nu}{C}$, where $C$ is the total number of configurations and 
$\nu \in \{1,2,\dots,C\}$. 

Figure~\ref{fig:rs-dyn_angle_kick_quench}b sketches the possible magnetic-field configurations 
($\mathbf{B}^{c1}, \dots, \mathbf{B}^{c4}$) for $C=4$ at a given lattice site. In this work,
we always choose site-randomly among the $C=4$ configurations. We checked that the results do
not depend on this choice. The TD Hamiltonian ensures that all relevant magnetic excitations 
are generated. Furthermore the linear-response regime is acquired for small $B$ 
and short $t_p$. A strong non-equilibrium regime is quickly obtained for large $B$ 
and large $t_p$. Note that local excitations are easily performed by setting $\mathbf{B}_{i}=0$
for the sites $i$ not to be excited.

To represent the magnetic excitations of the system in reciprocal space, the Fourier transform 
$\mexc$ of the TD magnetic moment $\mathbf{m} \left(\mathbf{r},t\right)$ is computed via
\begin{eqnarray}
\mexc \left(\mathbf{q},\omega\right) = \frac{1}{(2\pi)^3}  
\int d\mathbf{r} \int dt \; \mathbf{m} \left(\mathbf{r},t\right) 
{\rm e}^{-i \mathbf{q}\cdot\mathbf{r}} {\rm e}^{-i \omega t}, \label{eq:rs_dyn_mexc}
\end{eqnarray}
where $\mathbf{q}$ marks a point in the two-dimensional reciprocal space. The remaining
vector structure of $\mexc$ enables a separation of longitudinal ($x$-direction) and 
transverse ($y$- and $z$-direction) modes.
The real-space grid limits the resolution in $\mathbf{q}=(q_y,q_z)$ and leads to a certain
sampling of the Brillouin zone, shown in Fig.~\ref{fig:rs-dyn_angle_kick_quench}c for the
6$\times$6 lattice. Until stated otherwise, the time 
evolution captures times until a final propagation time $t_{\rm tot}$=1800. This 
provides a proper high-energy resolution of the magnetic excitations of the order of 
$\Delta\omega=0.003$.

\subsection{Slater-to-Heisenberg transition at half filling\label{slahe}}
\begin{figure}[t]
\hspace*{-0.2cm}\includegraphics*[width=8.9cm]{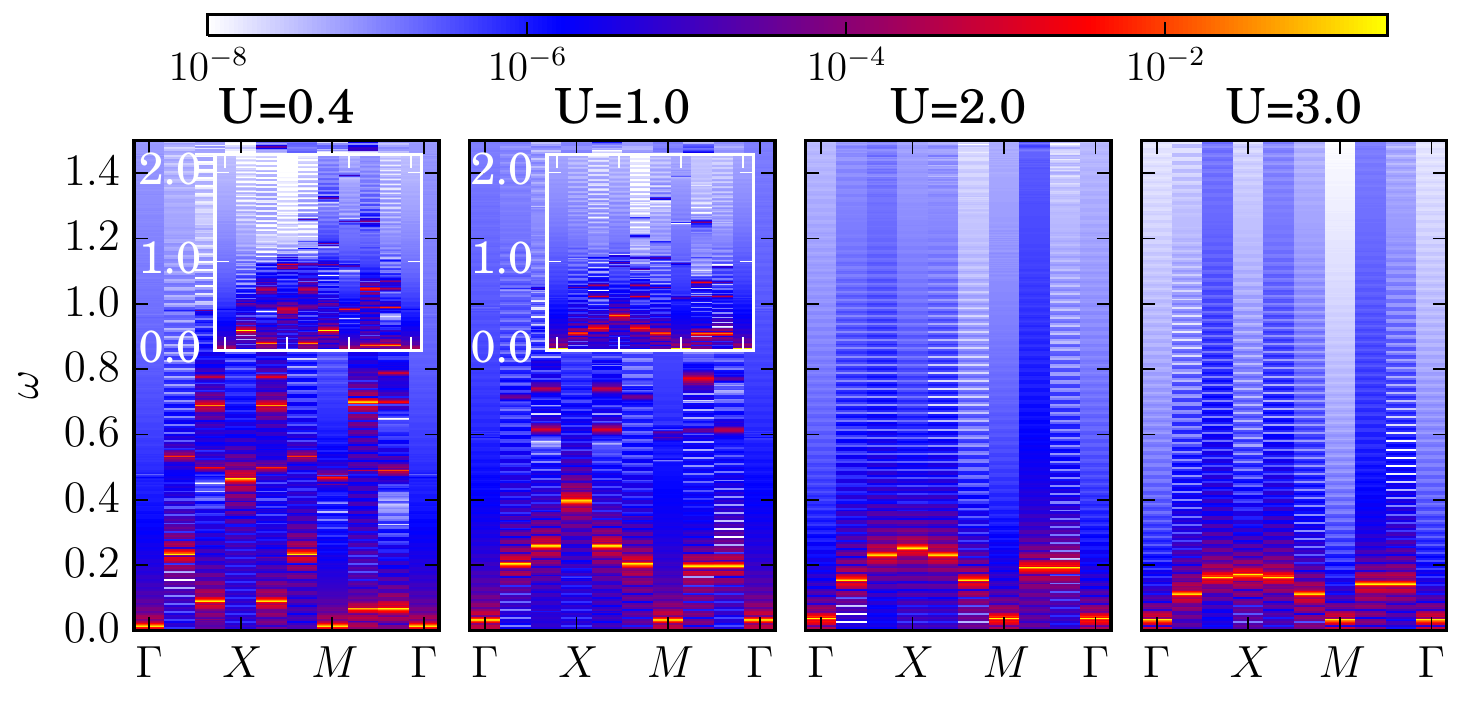}
\caption{(color online) $q$-dependent spin-excitation spectra along high-symmetry
lines witin the linear-response limit for different interaction strength $U$ at
half filling ($B=0.002$, $t_p=5$).}\label{fig:slaheis}
\end{figure}

The comparison between spin excitations in the Slater and in the Heisenberg limit of the 
Hubbard model at half filling serves as a first illustration. For small local interaction 
strength $U$ the system is close to a degenerate Fermi gas, with nearly absent local-moment 
physics. But spin polarization may set in via AFM order in reciprocal space
through the formation of Slater bands, separated by $U$. Stoner-like excitations amount to 
inter-band transitions and therefore give rise to a broad continuum of spin excitations,
formally starting off at $q=0$ with energy $U$. On the other hand in the large-$U$ limit 
with well-localized electrons, kinetic exchange $\sim$$t^2/U$ leads again to AFM order, now
between Heisenberg spins. Spin waves (or magnons) are the low-energy excitations, with
well-defined dispersion in $q$-space. Figure~\ref{fig:slaheis} documents this 
Slater-to-Heisenberg transition in the magnetic-excitation spectrum as obtained by  
real-space TD-RISB within linear response.
Though the finite number of $q$-points from the 36 lattice site limits the resolution, the
transformation of the broad excitation spectrum for small $U$ to the characteristic AFM 
magnon dispersion is evident. At $U=2$ this dispersion seems already formed, being
further renormalized at $U=3$ to the distinctive width $w=0.2$. 
\subsection{Finite doping at strong coupling}
\begin{figure}[t]
\hspace*{-0.2cm}\includegraphics*[width=8.9cm]{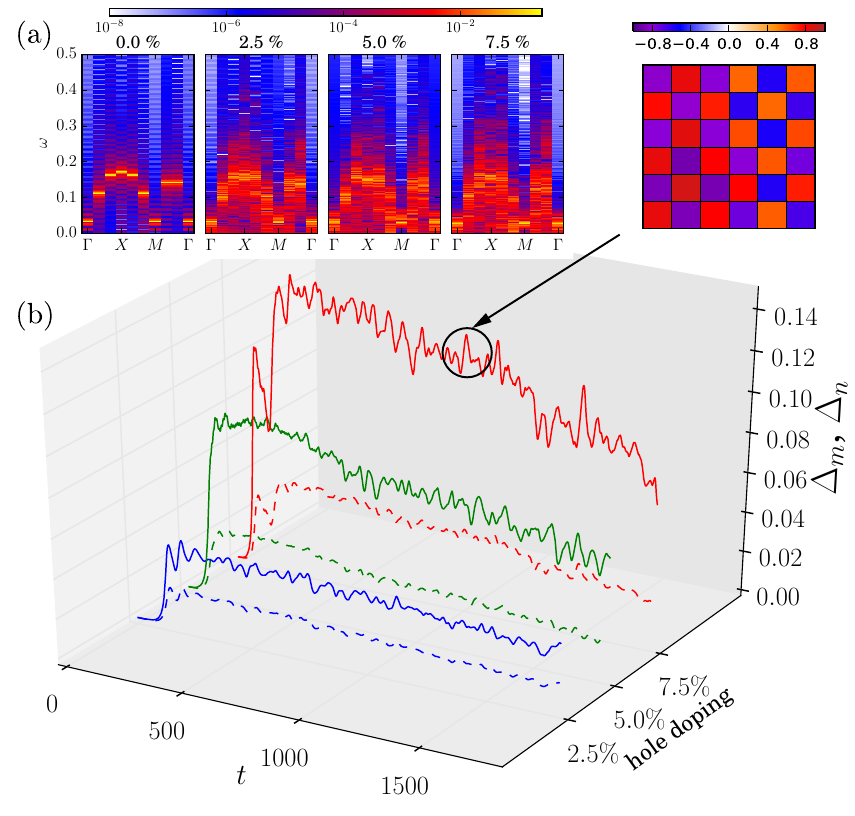}
\caption{(color online) Lattice excitation for $U=3$ for different dopings ($B=0.002$, $t_p=5$).
(a) $q$-dependent spin-excitation spectra along high-symmetry lines.
(b) Time evolution of the average inter-site charge (dashed lines) and 
spin (full lines) differences $\Delta_{m,n}$. At $t=1000$ the site-resolved magnetic
moment in real space is depicted.}\label{fig:dop}
\end{figure}

When introducing holes into the system, the lattice model becomes metallic at smallest 
doping. In $q$-space, the itinerant background is effective in substantially broadening the 
$U=3$ magnon spectrum and rendering it quickly incoherent (see Fig.~\ref{fig:dop}a).
To measure the real-space lattice amplitude fluctuations, it is instructive to define for a 
given local quantity $Q$ the average TD inter-site difference
\begin{equation}
\Delta_{Q}(t)=\frac{2}{N\,(N-1)}\sum_{i,j>i}\left|Q_i(t)-Q_j(t)\right|\quad.\label{eq:fluc}
\end{equation}
Figure~\ref{fig:dop}b displays $\Delta_Q(t)$ for the absolute value of the local moment
$m=|{\bf m}|$ and the local charge $n$, at finite hole doping after the short magnetic 
pulse ($t_p=5$). The inter-site differences grow with doping, i.e. increasing metallicity 
leads to stronger inter-site fluctuations, in line with the magnon destruction. As expected, 
while charge and spin differences seem to act rather coherently, the overall magnitude of 
$\Delta_{m}(t)$ is much larger. The strong coupling regime suppresses substantial charge 
fluctuations.

Especially shortly after the pulse, magnetic-moment fluctuations become large and decrease 
with $t$. Hence time-dependent fluctuations between lattice sites allow for an equilibration 
of the local observables. Since possible instabilities are encoded in linear-response functions,
the fluctuations are not random but are concerted such as to render phase-separating tendencies 
observable at certain propagation times (see real-space inset in Fig.~\ref{fig:dop}b). On the
other hand, with increasing doping of the strongly correlated AFM lattice, global excitations 
are truly effective in providing lattice disorder in the time domain. The decrease of 
the equilibrium AFM order parameter with doping can thus be used to drive a time-dependent 
phase separation into magnetic domains with different sizes of the staggered spin moment. This
loss of phase coherency is also part of the reduced magnon lifetime.

\subsection{Beyond linear response: Magnon destruction and revival} 
At half filling, the stability of the strong-coupling magnon dispersion is ensured by the 
linear-response limit of short pulse time $t_p$ and small magnetic-pulse fields. An 
adiabatic-like description, focussing on the slow dynamics of the stable local moments
(formed by the fast dynamics of the electronic degrees of freedom) is sound. However with
strong perturbations in magnetic field and pulse time, the degree of non-adiabticity is
expected to rise in charge-fluctuating Hubbard systems due to increased incoherent excitation 
of electrons. In other words, the identification of well-defined slow degrees of freedom and
their coherent modes may become difficult. A modeling focus within a sole adiabatic scheme may
be too restrictive. 
\begin{figure}[b]
\hspace*{-0.4cm}\includegraphics*[width=8.75cm]{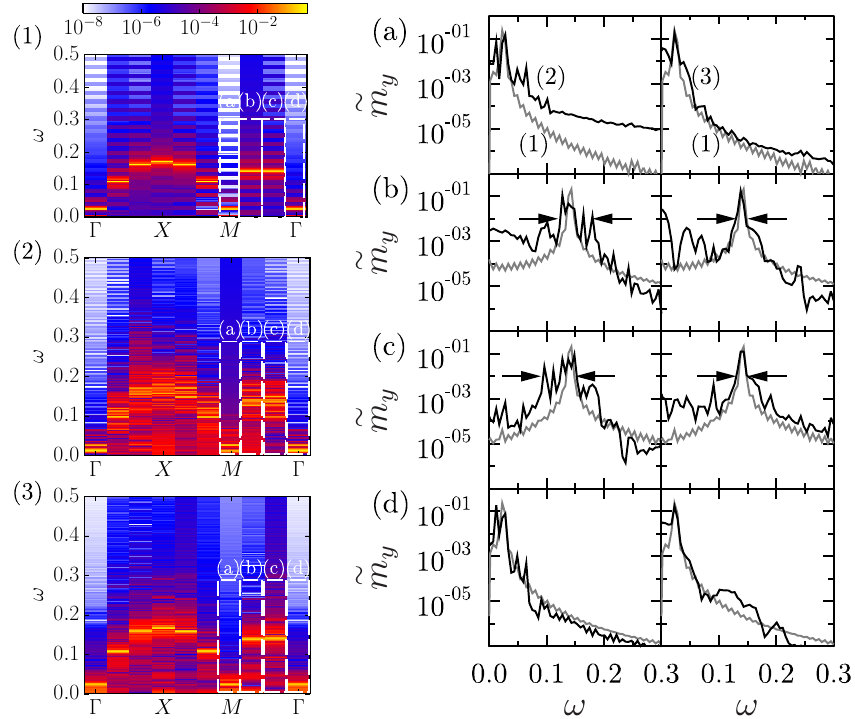}
\caption{(color online) $q$-dependent magnetic-excitation spectra $(U=3)$ with increasing
TD perturbations at half filling. Left: along high-symmetry lines for increasing pulse time 
$t_p=5, 80, 320$ in (1), (2) and (3). Right: spectral intensity in the $M-\Gamma$ 
direction (a-d), as given in the left part of the figure. Arrows indicate the peak sharpening
when comparing $t_p=80$ (2) with $t_p=320$ (3) in this direction within the Brillouin zone.
}\label{fig:magins}
\end{figure}

In Fig.~\ref{fig:magins} we provide results for such a non-linear regime of the dynamic Hubbard
lattice model. A sole increase of the absolute value of the pulsed magnetic field, while keeping
the pulse time short $(t_p=5)$ does not modify the AFM magnon dispersion by
clear means. But an increase of $t_p$ results in split-off sidebands to the dispersion 
away from ${\bf q}=\Gamma$. Stimulated by the higher magnetic field, a longer pulse time 
is effective in significantly reducing the magnon lifetime.
At $t_p=80$ the magnon dispersion appears destructed, only to surprisingly recover at 
$t_p=320$, especially inbetween the high-symmetry points $\Gamma$ and $M$. Such physics has 
already been observed by Zhitomirsky and  Chernyshev~\cite{zhi99} who 
studied magnons in the Heisenberg limit within the self-consistent Born approximation. There 
the magnons are destroyed above a critical-field strength due to an overlap of the 
single-magnon peak with the two-magnon continuum. A further increase of the magnetic field 
then lead to a reformation of the magnon spectrum. Hence the different energy transfers in 
our varying pulse times modify the magnon-magnon scattering such that long-lived spin waves 
become possible again at larger $t_p$. The quantitative details of our findings will depend
on the size of the simulated time window $t_{\rm tot}$, but our choice $t_{\rm tot}\gg t_p$ 
ensures the qualitative result.

\subsection{Beyond linear response: Local excitation in the doped AFM Mott state\label{locx}}
\begin{figure}[t]
\hspace*{-0.25cm}\includegraphics*[width=8.75cm]{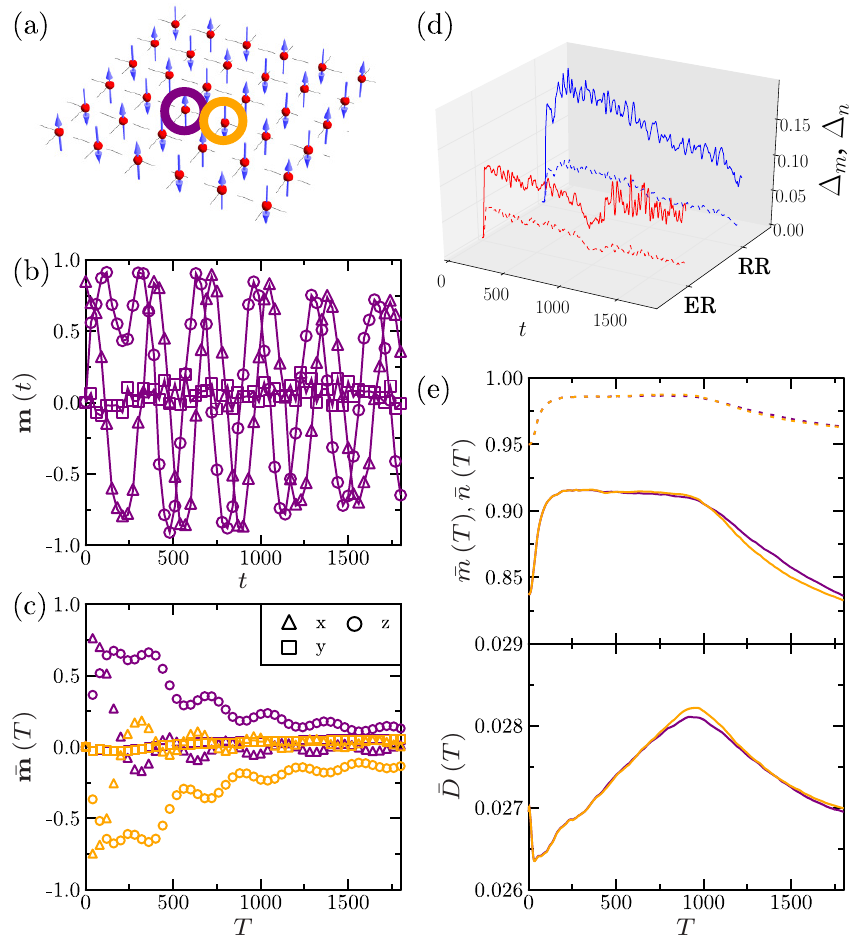}
\caption{(color online)
Local $\pm B_z$ excitation of two lattice sites in the 5\% hole-doped 
AFM state ($U=3$, $B=0.02$, $t_p=320$). 
(a) Lattice with excited sites encircled in purple(dark)/orange(grey). In (b),(c) and (e), the 
data in purple(dark) is associated with the first excited site, data in orange(grey) with the
second excited site as depicted in (a). (b) component-resolved TD local magnetic
moment ${\bf m}$ for one of the excited sites and (c) time-integrated local moment for both
excited sites. (d) TD lattice charge and 
spin fluctuations, resolved via the $\Delta_Q$-function (see text). (e) Time-integrated 
data on the respective excited sites for (top) the absolute value of the local magnetic 
moment $m$ and local charge $n$, as well as for (bottom) the local double occupation $D$.}
\label{fig:dflip}
\end{figure}

Finally, to show that our method is not bound to global excitations, a
local-excitation beyond linear response is studied. We start from a 5\% hole-doped 
antiferromagnetic state (staggered moment along $x$-direction) at strong coupling and apply 
an intense and long magnetic pulse ($B=0.02$, $t_p=320$) to two AFM-aligned adjacent 
lattice sites (see Fig.~\ref{fig:dflip}a). The pulsed magnetic field is directed along 
$+z$($-z$) on the first(second) site, to render spin momentum conserved by the excitation 
process. As shown in Fig.~\ref{fig:dflip}b, the $+z$-pulse-excited site switches its 
magnetic moment shortly after the pulse in that direction. But at the end of the pulse 
duration the moment points along $(-x,z)$. For longer times, the local moment oscillates
mainly in the $xz$-plane. The moment of the second site has a similar time evolution with 
proper sign changes. It is instructive to display the time-{\sl averaged} local quantities,
i.e. $\bar{Q}(T)=\frac{1}{T}\int_0^T dt\, Q(t)$. Notably for both local moments, time-averaging 
leads to a loss of a net resulting local magnetization axis (cf. Fig.~\ref{fig:dflip}c).

This paramagnetic behavior at long times holds not only for both excited sites, the whole 
lattice becomes disordered. Figure~\ref{fig:dflip}d  exhibits that the inter-site 
amplitude fluctuations differ if for $\Delta_{m,n}(t)$ the sum over sites in 
eq.~(\ref{eq:fluc}) is chosen as $i,j=$'excited,rest' (ER) or  $i,j=$'rest,rest' 
(RR). This means, that for ER only the fluctuations between the excited sites and the
remaining lattice sites enter $\Delta_{m,n}(t)$, while in the RR mode, fluctuations
between all sites but the excited ones are inspected.
While the charge fluctations are rather similar for both modes, the RR spin fluctuations are 
surpisingly stronger and monotonically decaying in time. On the other hand the ER spin 
fluctuations are smaller below a characterstic time $t_c\sim 1000$, i.e. there the excited 
spins exchange weaker with the other sites. After $t_c$ the fluctuations grow and finally
even become larger than in the RR sector. The time $t_c$ appears as a pulse-induced 
coherence time, since during $t_c$ the build-up larger local-moment amplitude and local 
charge on the both excited sites remain nearly constant (see Fig.~\ref{fig:dflip}e), 
forming a plateau-like structure. Only after $t_c$ the moment and charge relax to their 
respective values before the pulse. The time-averaged double occupation $\bar{D}(T)$ 
increases in the plateau, i.e. the local-correlation strength on the excited sites reduces 
somewhat below $t_c$. Note that the plateau formation results from the high-field pulse and 
seems not very sensitive to the pulse length $t_p$. This model local-excitation scenario 
documents the principle possibility of inducing/controlling local coherency in doped 
correlated magnets.
\section{Conclusions}
We presented a novel general and efficient TD-RISB formulation to deal 
with the problem of interacting lattice spins within a correlated electron system. 
The method is suitable in the limit of weak as well as strong correlations and can describe 
both, itinerant or Mott-insulating environments. Global and/or local excitations are handable,
with the resolution of $q$- as well as $r$-dependent out-of-equilibrium features. We chose 
some test cases to elucidate the general possibilities of the framework. Of course, the 
site-random magnetic-field pulse may not be linked to common experimental excitation 
protocols. It was here mainly utilized to illustrate the method. Implementation of realistic 
pump-probe processes is a next natural step. Also treating multi-orbital 
systems with spin-orbit coupling is highly interesting to resolve TD transfers 
between angular- and spin-momentum. Coupling to phonons would allow to account for
spin-lattice relaxations. An inclusion of explicit inter-site self-energies in cluster
extensions is furthermore possible. Finally, the approach is ideally suited to be
combined with TD density functional theory to advance on the description of realistic 
non-equilibrium physics.~\cite{kri15} It could overcome restrictions in the time-domain
modeling due to the presently used exchange-correlation functionals.

\begin{acknowledgments}
We thank C. Ederer and M. Sayad for helpful discussions.
This research was supported by the DFG-SFB925.
M.I.K. acknowledges support from European Research Council (ERC) Advanced Grant 
No. 338957 FEMTO/NANO. Computations were performed at the University of Hamburg 
and at the North-German Supercomputing Alliance (HLRN) under Grant No. hhp00026.
\end{acknowledgments}

\appendix*

\section{Dynamic one-band Mott transition}
As a reference, we consider the one-band Hubbard Hamiltonian and provide results for the 
time-dependent Mott problem after an interaction quench. Our data may be directly compared
to the one based on the original TD-Gutzwiller study of Schir{\'o} and 
Fabrizio.~\cite{schiro_2010} Here we employ the problem on a simple cubic lattice with 
nearest-neigbor dispersion of bandwidth $W_{\rm sc}=12\tau$. Note that for these calculations
we do not use a real-space approach, but define the primitive unit cell and solve the
problem by utilizing a fine-grid $k$-point mesh. In RISB, the equilibrium Mott 
transition occurs at $U_c=2.65\,\frac{W_{\rm sc}}{2}$. As a natural choice, one rescales
the energetics of the non-equilibrium problem in units of the critical equilibrium $U_c$.
The relevant interaction parameters are then expressed via $u_{\rm ini}=U_{\rm ini}/U_c$ 
and $u_{\rm fin}=U_{\rm fin}/U_c\equiv u$. 

\begin{figure}[t]
\hspace*{-0.1cm}\includegraphics*[width=8.75cm]{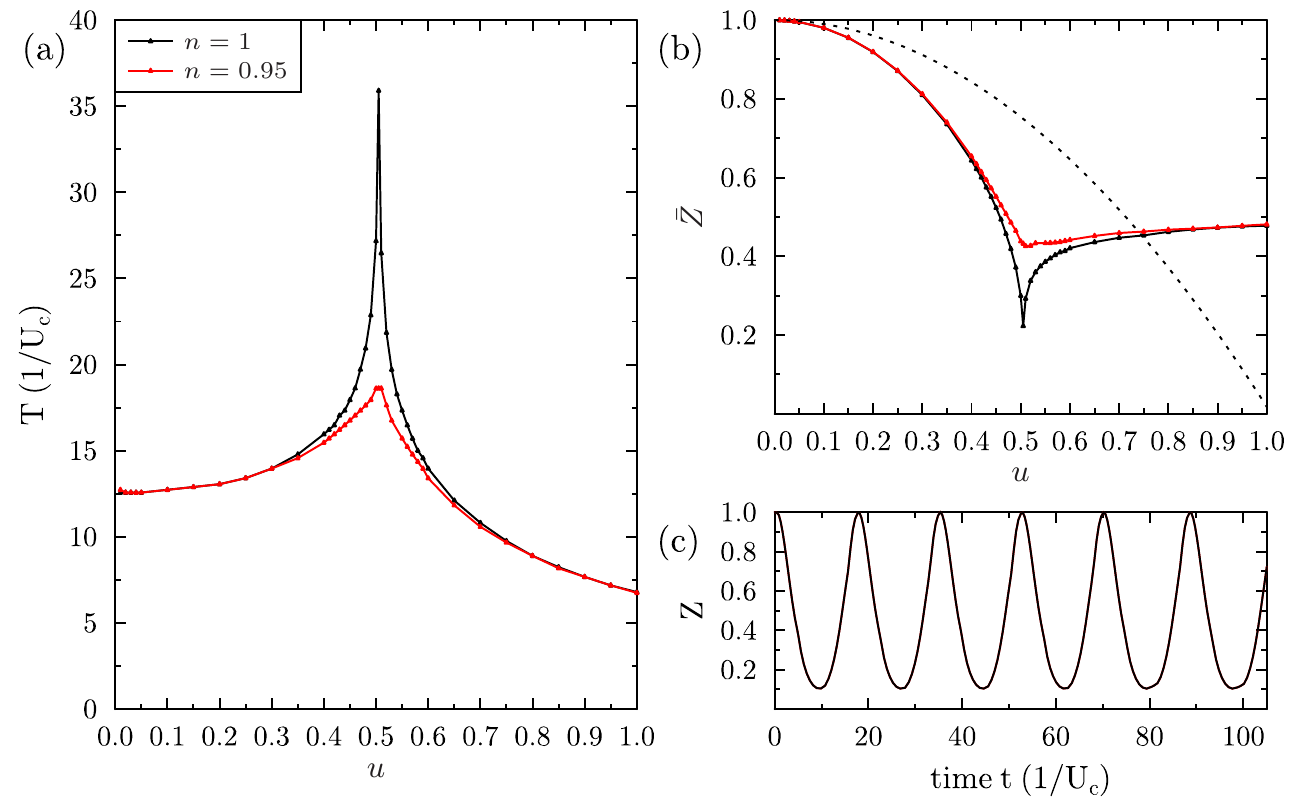}
\caption{(color online) TD-RISB data for the one-band dynamical Mott transition on a simple-cubic
lattice. (a) Period $T$ of oscillations of the QP weight $Z(t)$ with reduced quenched $u$
at half filling $n$=1 and at hole doping $n=0.95$.
(b) Time-averaged $\bar{Z}$ with $u$ in half-filled and doped case. Dashed line displays the
equilibrium $Z$.
(c) Time dependent $Z$ for the hole-doped case $n=0.95$ at $u$=0.5.
Note that the finite cut-off of the divergences in (a,b) at half filling is an 
numerical issue.}\label{fig:mit}
\end{figure}

From Ref.~\onlinecite{schiro_2010}, the dynamical critical interaction strength is given by
$u_{c}=(1+u_{\rm ini})/2$. Since TD-Gutzwiller and mean-field TD-RISB yield equivalent
physics, and we here choose to put the inital $u_{\rm ini}$ to zero, the dynamical Mott 
transition should occur at $u_c=0.5$ in our calculations. Indeed at half filling with this 
value of $u$, Fig.~\ref{fig:mit}a displays the characteristic logarithmic divergence of the 
period $T$ of oscillations in the QP weight $Z$ (for details see Ref.~\onlinecite{schiro_2010}).
Small hole doping is shown to cut-off this divergence. As shown in Fig.~\ref{fig:mit}b, the
time-averaged QP-weight $\bar{Z}=\frac{1}{t}\int_0^t dt'\, Z(t')$ has a similar signature as
the equilibrium $Z$, but in comparison the dynamical Mott transition takes place at half of the
equilibrium interaction strength. Finite doping prohibits a vanishing $\bar{Z}$ and hence the
dynamical Mott transition (as in the equilibirum case) remains absent in that case. 

For illustration, Fig.~\ref{fig:mit}c exhibits the oscillating $Z(t)$ in the hole-doped scenario. 
These oscillations are due to the mean-field description of the TD-Gutzwiller and TD-RISB method, 
which lack relaxation due to quantum fluctuations. However note that relaxation and/or thermalization
phenomena may still occur by other means through an increase of the numbers of degrees of freedom 
in the system under consideration. This was shown in recent TD multi-orbital 
studies~\cite{beh13,beh15} and is also effective in the present real-space work.

\bibliography{bibextra,bibmore}

\end{document}